\pdfoutput=1
\documentclass[11pt,a4paper]{article}
\usepackage[T1]{fontenc}
\usepackage[utf8]{inputenc}
\usepackage{lmodern}
\usepackage{microtype}
\usepackage{geometry}
\usepackage{amsmath,amssymb,bm}
\usepackage{graphicx}
\usepackage{cite}
\usepackage{xcolor}
\usepackage[colorlinks=true,linkcolor=blue,citecolor=blue,urlcolor=blue]{hyperref}

\allowdisplaybreaks

\newcommand{\dd}{\mathrm{d}}
\newcommand{\ii}{\mathrm{i}}

\newcommand{\psibar}{\overline{\psi}}
\newcommand{\qbar}{\overline{q}}
\newcommand{\Lag}{\mathcal{L}}

\title{Baryonic Bound States in the Non-Local NJL Model}

\author{Arpan Chatterjee\thanks{Talk presented at the 5th CERN Baltic Conference (CBC 2025), Kaunas University of Technology, Kaunas, Lithuania, 14--16 October 2025.} , Stefan Groote\\
\small Institute of Physics, University of Tartu, W. Ostwaldi 1, EE-50411 Tartu, Estonia\\
\small \texttt{arpan.chatterjee@ut.ee}, 
\small \texttt{stefan.groote@ut.ee}
}

\date{}

\begin{document}
\maketitle

\begin{abstract}
\noindent Baryons, as three-quark bound states, require a covariant treatment in the
intermediate-energy regime where perturbative QCD is no longer applicable and where nonperturbative correlations dominate.  This article reformulates the content of the CERN Baltic Conference 2025 presentation on baryonic bound states in the non-local Nambu--Jona-Lasinio (NJL) model. We review
how the relativistic Faddeev approach reduces the three-body quark problem to an effective quark--diquark bound-state problem, describe the scalar and axial-vector diquark channels, and show how the resulting quark--diquark Bethe--Salpeter
equation can be written as an eigenvalue problem for the baryon mass. The non-local NJL framework, motivated by QCD-based nonlocal interactions and
Dyson--Schwinger considerations, provides a compact description in which baryon masses and form factors are extracted from the numerical solution of coupled integral equations.
\end{abstract}

\section{Introduction}

Baryons are composite subatomic particles made of three quarks.  Compared with mesons, they involve a more intricate internal structure because the relativistic bound-state problem is genuinely a three-body problem. In the constituent-quark
picture this structure is classified by quark content and quantum numbers, while in a field-theoretic treatment one must also account for dressed propagators, correlations and nonperturbative effects.\\
A practical route is provided by the relativistic Faddeev approach~\cite{faddeev1993,ahlig2001,oettel2002,rezaeian2005}.  In this framework a baryon is treated as an effective bound state of a correlated quark pair, the diquark, and a third spectator quark. The diquark is not an asymptotic particle, but a useful dynamical correlation inside the baryon. This reduction converts the full three-quark problem into a quark--diquark Bethe--Salpeter equation while preserving covariance.\\
The purpose of this note is to summarise the formal pathway from the three-quark correlator to the Faddeev equation and then to the quark--diquark
Bethe--Salpeter representation relevant for  baryonic bound states in a non-local
NJL model~\cite{rezaeian2005,rezaeian2004,bowler1995,plant1998,diakonov1988}. The non-local NJL extension is important because nonlocality changes the structure of the effective quark interaction and is expected to lower the baryon ground-state mass relative to local models, indicating a more compact
state~\cite{rezaeian2005,rezaeian2004}.

\section{Relativistic Faddeev approach}

Following the relativistic three-body formulation of the Faddeev approach, the dominant three-quark amplitude may be written schematically as
\begin{equation}
  \Psi_{\alpha\beta\gamma}(p_1,p_2,p_3)
  = \langle 0 | T\{ q_\alpha(p_1) q_\beta(p_2) q_\gamma(p_3) \} | B(P) \rangle ,
  \label{eq:three_quark_amplitude}
\end{equation}
where $|B(P)\rangle$ is a baryon state with total momentum $P=p_1+p_2+p_3$. The object in Eq.~\eqref{eq:three_quark_amplitude} is to be
understood as a dressed amplitude: in principle, it contains nontrivial vacuum structure and therefore includes effects of condensates, sea quarks and gluons.

A convenient starting point, used in covariant quark--diquark reductions of the baryon problem~\cite{ahlig2001,oettel2002}, is the six-point function
\begin{equation}
  G(\{x_i\},\{y_i\})
  = \langle 0 | T\prod_{i=1}^{3} q(x_i)\,\qbar(y_i) |0\rangle .
  \label{eq:six_point}
\end{equation}
In momentum space, a baryonic bound state of mass $M$ appears as a pole of this
correlator,
\begin{equation}
  G(\{p_i\},\{q_i\}) \approx
  \frac{\ii\,\Psi(p_1,p_2,p_3)\,\overline{\Psi}(q_1,q_2,q_3)}
       {P^2-M^2} ,
  \label{eq:pole_ansatz}
\end{equation}
where, $P=p_1+p_2+p_3=q_1+q_2+q_3$, is
the total momentum of the 3 quark baryonic state expressed by incoming momenta $q_i$ and outgoing momenta $p_i$.

Let $G_0$ denote the disconnected product of three fully dressed quark propagators and let $K$ denote the three-quark scattering kernel, including two-
and three-particle irreducible contributions.  In the standard covariant bound-state construction~\cite{oettel2002}, the Dyson equation for the six-point function is
\begin{equation}
  G = G_0 + G_0 \circ K \circ G ,
  \label{eq:dyson}
\end{equation}
where $\circ$ denotes the convolution over internal momenta and discrete indices.
Inserting the pole form, Eq.~\eqref{eq:pole_ansatz}, into
Eq.~\eqref{eq:dyson} and comparing residues yields the homogeneous bound-state equation
\begin{equation}
  \Psi = G_0 \circ K \circ \Psi .
  \label{eq:homogeneous}
\end{equation}
\begin{equation}
  \lower18pt\hbox{\includegraphics[scale=0.3]{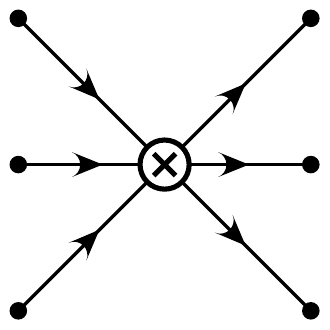}}\ =\
  \lower18pt\hbox{\includegraphics[scale=0.3]{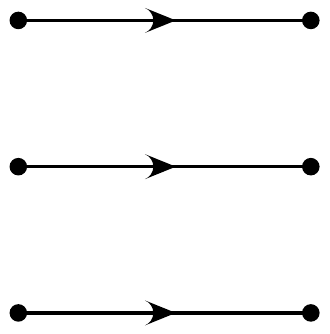}}\ +\
  \lower24pt\hbox{\includegraphics[scale=0.3]{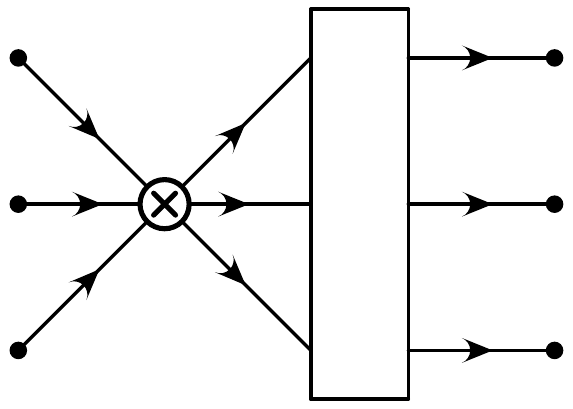}}
\end{equation}
Equivalently,
\begin{equation}
  G^{-1}\circ \Psi =0,
  \label{eq:dirac_baryon}
\end{equation}
which may be regarded as the "Dirac-type" equation for the baryonic state.

\section{Faddeev approximation}

Equation~\eqref{eq:homogeneous} is not directly solvable because the exact kernel $K$ is unknown.  In the Faddeev approximation~\cite{oettel2002}, irreducible three-quark interactions are neglected and the kernel is decomposed into pairwise interactions,
\begin{equation}
  K = K_1 + K_2 + K_3,
  \label{eq:kernel_sum}
\end{equation}
where $K_i$ describes the interaction of the two quarks other than $i$, and the index $i$ labels the spectator quark.

The full amplitude is decomposed into three Faddeev components,
\begin{equation}
  \Psi = \Psi_1+\Psi_2+\Psi_3,
  \qquad
  \Psi_i = G_0\circ K_i\circ \Psi .
  \label{eq:faddeev_components}
\end{equation}
After isolating the two-quark $t$-matrix in the relevant channel, the components can be written in terms of diquark correlations, following the separable two-quark approximation used in relativistic baryon calculations~\cite{oettel2002}.  With $D_i$ denoting the diquark correlation in the pair opposite the spectator quark $i$, one obtains schematically,
\begin{equation}
  \Psi_i = G_0 \circ D_i \circ (\Psi_j+\Psi_k),
  \qquad (i,j,k)\; \text{cyclic} .
  \label{eq:faddeev_diquark}
\end{equation}
Thus, the baryon is described through repeated exchange of quarks between diquark--quark configurations. The Faddeev Equation (\ref{eq:faddeev_diquark}) can then be written as:
        \begin{align}
            \phi^a_{i,\alpha} = \int \chi^{a}_{i,\beta \gamma} S^{\beta \beta'}_j S^{\gamma \gamma'}_k \sum_{b,b'} \Bar{\chi}^b_{j,\gamma',\alpha} d^{(j)}_{bb'}\phi^{b'}_{j,\beta'} + (j \leftrightarrow k)
        \end{align}
        \begin{figure}[h]
        \centering
        \begin{minipage}[c]{0.2\textwidth}
            \centering
            \includegraphics[width=\linewidth]{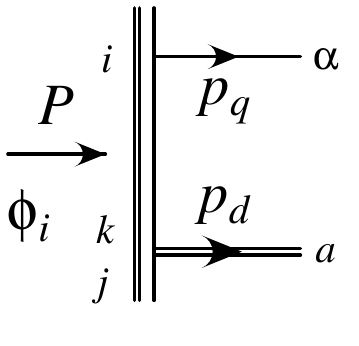}
        \end{minipage}
        \begin{minipage}[c]{0.05\textwidth}
            \centering
            { $=$}
        \end{minipage}
        \begin{minipage}[c]{0.27\textwidth}
            \centering
            \includegraphics[width=\linewidth]{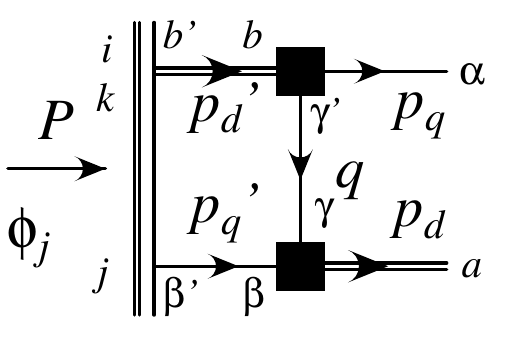}
        \end{minipage}
        \makebox[0pt][l]{\hspace{0.5em}{+ $(j \leftrightarrow k)$}}
        \end{figure}

\section{Kinematics}

For a chosen spectator-quark channel the total momentum is decomposed as
\begin{equation}
  P=p_1+p_2+p_3=p_i+p_j+p_k=p_q+P_d,
  \label{eq:total_momentum}
\end{equation}
where $p_q$ is the spectator-quark momentum and $P_d$ is the diquark momentum.
Introducing a momentum partitioning parameter $\eta\in[0,1]$, as commonly used in the quark--diquark Bethe--Salpeter formulation, the relative
momentum $p$ is
\begin{equation}
  p=(1-\eta)p_q-\eta P_d=(1-\eta)p_i-\eta(p_j+p_k)=p_i-\eta P .
  \label{eq:relative_momentum}
\end{equation}
Equivalently,
\begin{equation}
  p_q=\eta P+p,
  \qquad
  P_d=(1-\eta)P-p .
  \label{eq:q_d_momenta}
\end{equation}
The exchanged-quark momentum appearing in the quark-exchange kernel is
\begin{equation}
  q= -p-p' + (1-2\eta)P .
  \label{eq:exchanged_momentum}
\end{equation}
Physical observables such as the baryon mass and form factors must be independent of the unphysical parameter $\eta$; in numerical work this condition
is used as a stability criterion.

\section{Diquark channels}

As a minimal set for describing octet and decuplet baryons, one retains scalar and axial-vector diquark channels~\cite{oettel2002,rezaeian2005,rezaeian2004}. The two-quark correlation in channel $i$ is written as
\begin{align}
  D_i(p_j,p_k;q_j,q_k)S_i(p_i)
  ={}& \chi_i^{[5]}(p_j,p_k)d_5(p_j+p_k)\overline{\chi}_i^{[5]}(q_j,q_k)
      \nonumber\\
  &+\chi_i^{[\mu]}(p_j,p_k)d_{\mu\nu}(p_j+p_k)
      \overline{\chi}_i^{[\nu]}(q_j,q_k),
  \label{eq:diquark_decomposition}
\end{align}
where $\chi$ and $\overline{\chi}$ are diquark vertex functions. The scalar and axial-vector propagators are taken in the forms
\begin{equation}
  d_5(P)=\frac{-\ii}{P^2-m_S^2+\ii\epsilon},
  \label{eq:scalar_propagator}
\end{equation}
and
\begin{equation}
  d_{\mu\nu}(P)=\frac{-\ii}{P^2-m_A^2+\ii\epsilon}
  \left(g_{\mu\nu}-\frac{(1-\xi)P_\mu P_\nu}{P^2-\xi m_A^2}\right),
  \label{eq:axial_propagator}
\end{equation}
where $m_S$ and $m_A$ are the scalar and axial-vector diquark masses.

\section{From Faddeev to Bethe--Salpeter form}

The quark--diquark wave function is obtained by attaching the spectator-quark and diquark propagators to the vertex function.  For channel $j$,
\begin{equation}
  \Psi^b_{j,\beta}(p,P)
  := S_{j,\beta\beta'}(\eta P+p)
     d^{(j)}_{bb'}\bigl((1-\eta)P-p\bigr)
     \Phi^{b'}_{j,\beta'}(p,P) .
  \label{eq:wavefunction_vertex}
\end{equation}
The Faddeev equation can then be recast as a quark--diquark Bethe--Salpeter equation,
\begin{equation}
  \Phi^a_{i,\alpha}(p,P)
  = \int\frac{\dd^4 p'}{(2\pi)^4}\,
  K^{ab}_{ij,\alpha\beta}(p,p',P)\,
  \Psi^b_{j,\beta}(p',P),
  \label{eq:bse_vertex}
\end{equation}
with exchange kernel
\begin{equation}
  K^{ab}_{ij,\alpha\beta}(p,p',P)
  := \chi^a_{i,\beta\gamma}(p',q)
     S_{k}^{\gamma\gamma'}(q)
     \overline{\chi}^{b}_{j,\gamma'\alpha}(q,p),
  \label{eq:exchange_kernel}
\end{equation}
where $q=(1-2\eta)P-p-p'$, is the momentum of the exchanged quark.\\
For a fixed value of $P^2$, the homogeneous equation can be written as an eigenvalue problem, as in numerical quark--diquark treatments of baryons~\cite{oettel2002},
\begin{equation}
  \lambda(P^2)\Phi^a(p,P)
  =\int\frac{\dd^4 p'}{(2\pi)^4}\,
  K^{ab}(p,p',P)\Psi^b(p',P).
  \label{eq:eigenvalue}
\end{equation}
The mass of the lowest bound state is determined by
\begin{equation}
  \lambda(M^2)=1 .
  \label{eq:mass_condition}
\end{equation}
In practical Euclidean calculations the corresponding sign convention for $P^2$ must be chosen consistently.

\section{Covariant expansion and numerical strategy}

The scalar and axial-vector components of the vertex and wave function may be expanded in a finite covariant basis. With $a=\{5,\mu\}$ one writes schematically
\begin{equation}
  \begin{pmatrix}
    \Phi^5(p,P) \\
    \Phi^\mu(p,P)
  \end{pmatrix}
  =
  \begin{pmatrix}
    \displaystyle \sum_{i=1}^{2} S_i(p,P)Y_i(p,P) \\
    \displaystyle \sum_{i=1}^{6} \gamma_5 A_i^\mu(p,P)Y_{i+2}(p,P)
  \end{pmatrix},
  \label{eq:phi_basis}
\end{equation}
and analogously for $\Psi$ with scalar functions $\widehat{Y}_i$.
\begin{align}
    \begin{pmatrix}
    \Psi^{5}(p, P) \\
    \Psi^{\mu}(p, P)
    \end{pmatrix}
        =
    \begin{pmatrix}
    \sum_{i=1}^{2} \mathcal{S}_{i}(p, P) \hat{Y}_{i}(p,P)  \\
    \sum_{i=1}^{6}  \gamma_{5} \mathcal{A}_{i}^{\mu}(p, P) \hat{Y}_{i+2}(p,P)
    \end{pmatrix} \hspace{1mm}
\end{align}
Once the baryon mass fixes $P^2=M^2$, the functions depend on $p^2$ and the angle
$z=\hat{p}\cdot \hat{P}$.\\
The numerical solution proceeds by expanding the angular dependence in Chebyshev polynomials and reducing the problem to coupled one-dimensional
integral equations for the Chebyshev moments of the scalar coefficient functions. The relation between vertex and wave-function coefficients can be written as
\begin{equation}
  \widehat{Y}_i=(g_0')_{ij}(p^2,z)Y_j(p^2,z),
  \label{eq:yhat_relation}
\end{equation}
while the eigenvalue equation becomes
\begin{equation}
  \lambda\,Y_i(p^2,z)
  = \int\frac{\dd^4 p'}{(2\pi)^4}
  (H')_{ij}(p^2,p'^2,z,z')\widehat{Y}_j(p'^2,z').
  \label{eq:scalar_integral_eq}
\end{equation}
The bound-state mass and form factors are then extracted from the converged solution, with independence from the partition parameter $\eta$ serving as a check of the calculation~\cite{ahlig2001,oettel2002}.

\section{Non-local NJL interaction}

The NJL model provides an effective four-fermion description of low-energy QCD. Non-local extensions of the NJL model have been developed to incorporate momentum-dependent interactions and quark confinement effects~\cite{rezaeian2005,bowler1995,plant1998,rezaeian2004,diakonov1988}. For the QCD-motivated non-local extension, one may motivate the interaction by starting from the QCD Lagrangian
\begin{equation}
  \Lag_{\mathrm{QCD}}
  = \psibar(x)\bigl(\ii\gamma^\mu\partial_\mu-m_q
     +g\gamma^\mu A^a_\mu T^a\bigr)\psi(x)
    -\frac{1}{4}F^a_{\mu\nu}F_a^{\mu\nu} .
  \label{eq:qcd_lagrangian}
\end{equation}
The gluon field equation can be represented through a nonlocal kernel
$G(x-y)$ as
\begin{equation}
  A_a^\nu(x)=\int \dd^4y\,G(x-y)\,g\psibar(y)\gamma^\nu T^a\psi(y),
  \label{eq:gluon_solution}
\end{equation}
where the Green's function satisfies
\begin{equation}
  (\partial_\mu\partial^\mu-M_g^2)G(x-y)=\delta^{(4)}(x-y).
  \label{eq:greens_function}
\end{equation}
Substitution into Eq.~\eqref{eq:qcd_lagrangian} gives the effective non-local NJL Lagrangian~\cite{frasca2022}
\begin{align}
  \Lag_{\mathrm{NJL}}
  ={}& \psibar(x)(\ii\gamma^\mu\partial_\mu-m_q)\psi(x) \nonumber\\
  &+g^2\psibar(x)\gamma^\mu T^a\psi(x)
  \int\dd^4y\,G(x-y)\,
  \psibar(y)\gamma_\mu T^a\psi(y) .
  \label{eq:nonlocal_njl}
\end{align}
This nonlocality modifies the diquark correlations entering the Faddeev kernel and is expected to affect the extracted baryon spectrum~\cite{bowler1995,plant1998,rezaeian2004}.  In particular, the non-local NJL calculation can lead to smaller baryon masses than the local
model, consistent with a more compact baryonic state~\cite{rezaeian2004}.

\section{Conclusions and outlook}

The relativistic Faddeev approach provides a covariant framework for describing baryons as quark--diquark bound states. Starting from the six-point correlator, the baryon appears as a pole, and the homogeneous bound-state equation follows
from the Dyson equation. By retaining scalar and axial-vector diquark channels, the three-body problem is reduced to a quark--diquark Bethe--Salpeter equation. The eigenvalue condition $\lambda(M^2)=1$ determines the baryon mass, while the covariant expansion and Chebyshev approximation provide a practical numerical
route for solving the coupled integral equations.

Future work should focus on implementing the non-local NJL interaction directly in the quark--diquark kernel, testing stability with respect to the partition parameter $\eta$, and comparing the resulting baryon spectrum and form factors
against both local NJL calculations and phenomenological expectations. This program is a step toward a nonperturbative description of baryon confinement, ground-state mass lowering and the internal structure of compact baryonic states.

\section*{Acknowledgements}

The first author thanks the organisers of the 5th CERN Baltic Conference and the Institute of Physics, University of Tartu.  This manuscript is based on material presented at CBC 2025.

\end{document}